\documentclass[sigconf,screen,]{acmart}

\setcopyright{rightsretained}
\copyrightyear{2020}
\acmYear{2020}

\acmBooktitle{Preprint}

\usepackage{cleveref}
\usepackage{eurosym}
\usepackage{color, colortbl}
\include{footnote}

\definecolor{Gray}{gray}{0.9}
\definecolor{grey}{rgb}{0.22, 0.22, 0.22}
\newcommand\usecase[1]{{\textit{#1}}}
\newcommand\navi{\usecase{Navigation}}
\newcommand\proa{\usecase{Proactivity}}
\newcommand\fami{\usecase{Family}}
\newcommand\cont{\usecase{Control}}
\newcommand\shar{\usecase{Sharing}}
\newcommand\tran{\usecase{Transparency}}

\newcommand\quot[2]{``#1''\emph{\small{-#2}}}

\begin{document}

\title[What If Your Car Would Care?]{What If Your Car Would Care? Exploring Use Cases For Affective Automotive User Interfaces}

\author{Michael Braun}
\affiliation{%
  \institution{BMW Group Research, New Technologies, Innovations}
  \institution{LMU Munich}
  \city{Munich}
  \country{Germany}}
\email{michael.bf.braun@bmw.de}

\author{Jingyi Li}
\affiliation{%
  \institution{LMU Munich}
  \city{Munich}
  \country{Germany}}
\email{jingyi.li@ifi.lmu.de}

\author{Florian Weber}
\affiliation{%
  \institution{BMW Group Research, New Technologies, Innovations}
  \city{Munich}
  \country{Germany}}
\email{florian.ww.weber@bmw.de}

\author{Bastian Pfleging}
\affiliation{%
  \institution{TU Eindhoven}
  \city{Eindhoven}
  \country{Netherlands}}
\email{b.pfleging@tue.nl}

\author{Andreas Butz}
\affiliation{%
  \institution{LMU Munich}
  \city{Munich}
  \country{Germany}}
\email{butz@ifi.lmu.de}

\author{Florian Alt}
\affiliation{%
  \institution{Bundeswehr University}
  \institution{LMU Munich}
  \city{Munich}
  \country{Germany}}
\email{florian.alt@unibw.de}

\renewcommand{\shortauthors}{M. Braun et al.}


\begin{abstract}
In this paper we present use cases for affective user interfaces (UIs) in cars and how they are perceived by potential users in China and Germany. Emotion-aware interaction is enabled by the improvement of ubiquitous sensing methods and provides potential benefits for both traffic safety and personal well-being.
To promote the adoption of affective interaction at an international scale, we developed 20 mobile in-car use cases through an inter-cultural design approach and evaluated them with 65 drivers in Germany and China. 
Our data shows perceived benefits in specific areas of pragmatic quality as well as cultural differences, especially for socially interactive use cases. We also discuss general implications for future affective automotive UI. 
Our results provide a perspective on cultural peculiarities and a concrete starting point for practitioners and researchers working on emotion-aware interfaces.
\end{abstract}


\maketitle


\section{Introduction}

Recent advances in machine learning and ubiquitous sensing technologies have spawned a trend to enhance interaction with empathic features. Especially designers of natural user interfaces, such as voice assistants, are starting to integrate user state detection to make their interfaces appear more human. One of the main lines of research working on automatic state recognition is \emph{Affective Computing}. Its main goal is to ``sense, interpret, adapt, and potentially respond appropriately to human emotions''~\cite{McDuff2018}, and this is usually realized by analyzing psycho-physiological features, speech, or facial expressions~\cite{Weber2018}. 
In addition, researchers in the automotive field have put their hopes in using affective systems to increase traffic safety, for example by preventing negative emotional states, which are statistically correlated with unsafe driving behavior~\cite{Jeon2014}. One viable technique that has been shown to regulate driver emotions is empathic voice interaction~\cite{braun2019strategies}, providing another incentive for the design of naturalistic UIs. 

\begin{figure}
  \centering
  \includegraphics[width=\columnwidth]{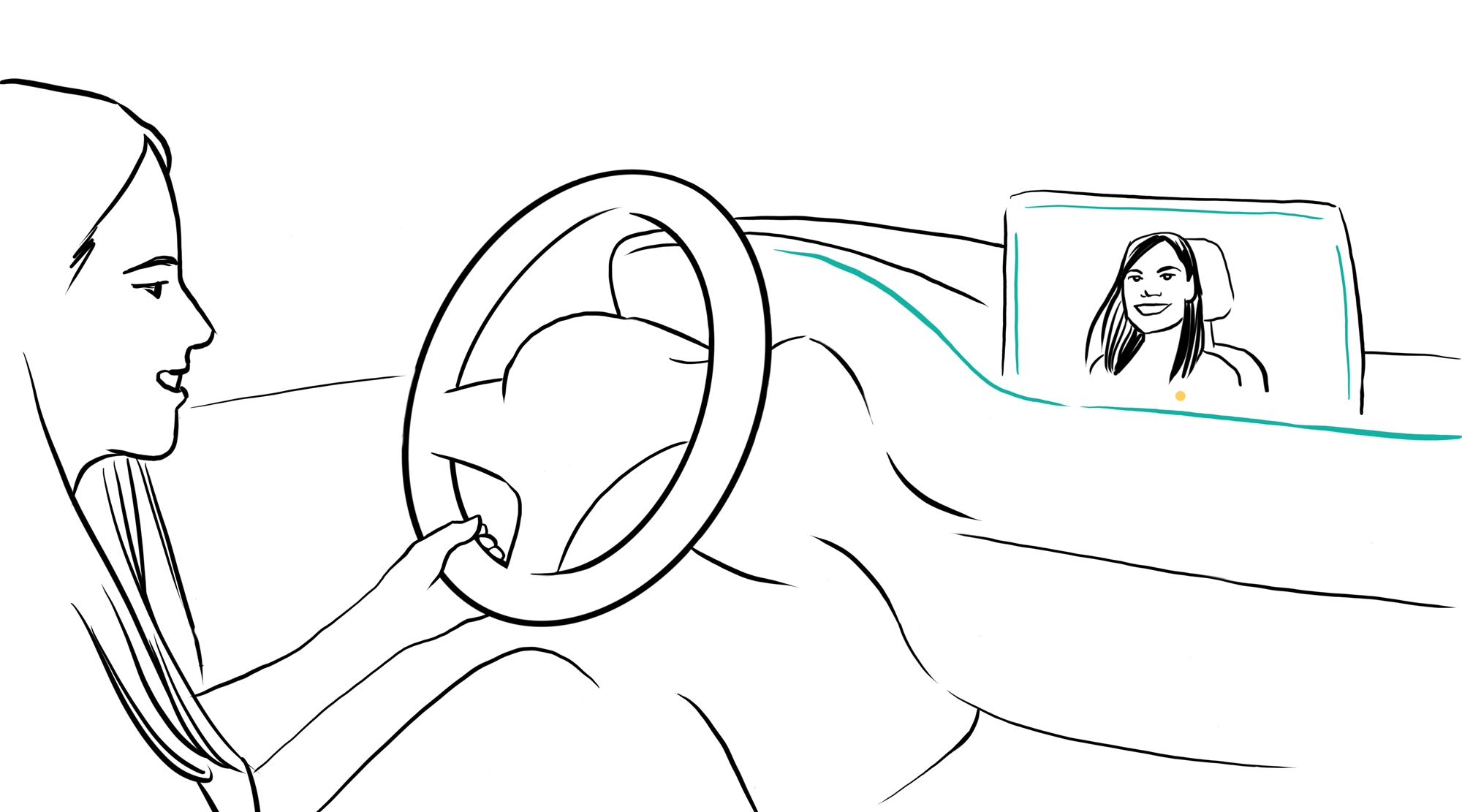}
  \caption{Drivers experienced 20 affective features implemented using camera-based facial expression analysis in in-car and mobile prototypes.}~\label{fig:teaser}
   \vspace{-0.6cm}
\end{figure}

While the ultimate goal of many affective systems is to understand users' emotions in order to perfectly tailor user interfaces to their expectations and needs, we recognize the potential of initially creating more basic use cases of emotion detection, in order to establish a raison d'\^etre and trust in the technology in the minds of customers. 
Only when a significant number of pioneers can be convinced of the benefits of affective systems, we can justify the efforts to launch emotion detection into the mainstream and so pave the way for more sophisticated systems in order to explore more wide-ranging research questions.
This challenge is made more difficult by the global segmentation of the car market, which forces manufacturers to stay flexible for the requirements of culturally divergent regions.

The use cases we present are designed to convince early adopters around the globe of the benefits of emotion-related interaction, which includes but is not limited to emotion-aware features. We want to spur UI designers from all disciplines to direct their efforts towards enabling affective systems, and keep cultural diversity in mind. Consequently, the use cases we present are derived and tested with users from two traditionally disparate cultures. We apply the ideas to the context of automotive UI for users from the currently globally largest car market of East Asia~\cite{wang2015fluctuation} and a classical car nation in Europe, by comparing feedback of Chinese and German users.

\subsection{Contribution}

We designed and implemented 20 use cases for emotion detection in the car, which we evaluated with 65 drivers in China and Germany. Users prefer affective approaches to enhance pragmatic features and proactive interaction fuelled by emotion detection. Findings are condensed into implications for future affective automotive user interfaces, regarding data privacy, cultural, and social responsiveness, and a reflection on how to make affective UIs accessible to the public.



\section{Related Work}

For this work, two areas of research are important: 
Emotion detection algorithms and applications of affective user interfaces form the technological basis for our research. 
In addition, we rely on the work of social scientists, who have set a foundation for understanding cultural backgrounds and how they evolve along with digitization and we focus on the automotive domain as exemplary application area.

\subsection{Affective Automotive User Interfaces}
Emotions have long been known to affect human behavior in general, which specifically also impacts road safety. Manifestations of both negative and positive emotional states, such as happiness and anger, have been found to degrade driving performance~\cite{Jeon2014}.
One possible way of counteracting emotion-related hazardous situations are affective user interfaces, which sense the driver's state and intervene when possibly dangerous behavior is detected. A first concept for this strategy was presented by Nasoz et al.~in 2002~\cite{Nasoz2002}. Since then, many concepts explored emotion-based in-car UIs to improve the drivers' emotions, mostly in the name of safety but also to improve hedonic factors and user experience (UX). Ideas range from micro-entertainments during breaks, e.g., at stop lights, to refocus the driver's attention towards the road~\cite{Alt2010} to empathic speech assistants, downplaying the mistakes of other drivers in order to avoid agitation over traffic~\cite{Nass2005}. Harris \& Nass found that a voice assistant which reframes frustrating events so that drivers see them in a more positive light can indeed lead to better driving behavior and less negative emotions~\cite{Harris2011}.

With technical advancements, more and more elaborate emotionally sentient digital assistants will become feasible~\cite{McDuff2018}.
Other applications of affective in-car UIs have been realized by giving direct feedback to the driver regarding their current state, so they can reflect and adapt if necessary~\cite{braun2019driverstate,Voelkel2018}; by actively influencing the driver, e.g., through music playback~\cite{Fakhrhosseini2014}, and, more subliminally, through ambient lighting~\cite{Hassib2019}; or by enhancing the driving experience in an emotional way, e.g., with emotional interaction in a feedback channel among nearby cars~\cite{Wang2016}.
Coughlin et al. envision an aware car which detects the driver's state from psycho-physiological data and influences them to stay in an optimum driving state~\cite{Coughlin2011}. The Yerkes-Dodson Law serves as  psychological ground work for their concept, which states that an optimum in performance is usually achieved at a medium level of arousal~\cite{Yerkes1908}.
Braun \& Alt extend the short-term reactiveness of this approach with a model that includes temporary states and permanent traits such as personality, driving experience and cultural affiliation~\cite{braun2019model}.

Emotion detection algorithms, based on EEG, GSR, ECG, respiration, body posture, facial expressions, or speech are presented in related literature. For the use in automotive environments, camera-based facial expression analysis has been established as a minimium viable sensor setup due to its non-contact application and low intrusiveness~\cite{Weber2018}. 
While research on the sensing of affective states is well advanced, applications from the literature often rely on imaginary systems providing highly sophisticated context information and naturalistic output modalities realizable only in labs. We agree that such futuristic concepts are important for the vision of affective UIs, but they also create a feasibility gap which complicates commercial progress. We close this gap by introducing feasible use case scenarios for emotional interaction in the car, based on a contemporary technology stack.


\subsection{Cultural Influences On Interaction}


In the global automotive industry, designing a universal UI achieving the same level of UX and satisfaction in different user groups, has always been a large challenge~\cite{heimgartner2017cultural}. Especially when designing emotional interfaces for one or more market segments, how enjoyable they are perceived to be varies considerably across cultures~\cite{rogers2011interaction}.
Previous studies investigating cultural differences in the automotive context found varying behavioural tendencies that influenced usability~\cite{khan2014study} as well as disparate design preferences for in-car HMIs between European and Asian users~\cite{khan2016cross}. 
Consequently, UI designers are increasingly creating interfaces with culture in mind, following a culturally sensitive design approach~\cite{lachner2019user,lachner2018culturally}. For example, Wang et al. conceived a traffic system, tested it with Swedish and Chinese drivers, and found identical requirements for simple traffic scenarios, whereas in complex traffic scenarios the different cultural backgrounds required different types of information~\cite{Wang2016same}. 
With Hofstede's cultural dimensions theory they build upon a traditional model to comprehend influences in UI design, based on cross-cultural comparisons for the dimensions power distance, individualism, masculinity, uncertainty avoidance, long-term orientation, and indulgence~\cite{del1996international,hofstede2010}.

In our work, German and Chinese drivers are considered distinct user groups, which can provide more diverse insights on emotional interfaces in two highly dissimilar cultures. The two countries, as the prevalent economies in two of the biggest premium car markets~\cite{sha2013upward}, are also meant to ensure a certain level of technology affinity in the car owners, whom we anticipate to be early adopters of novel in-car interaction like emotion detection features. According to Hofstede's work, Germany and China provide quite opposite profiles for individualism (G67/C20), power distance (G35/C80) and uncertainty avoidance (G65/C30)~\cite{hofstede2010}.
We thus expect requirements for enjoyable user interfaces to differ between both countries. Related work tells us that Chinese consumers set a higher value on appearances at first impressions, they value family time, and rather withhold their emotions to seem positive, while Germans appreciate sheer functionality and aim for personal satisfaction and individual solutions~\cite{Lachner2015}.
They also differ in terms of emotional engagement, driving skills, and traffic perception~\cite{chu2019traffic}, which is of course also contingent on the different traffic rules in place, such as the right of way, speed limits, or the severity of penalties~\cite{Li2019,zhang2018traffic}. Chinese drivers also show different driving preferences, styles and behaviors from the German, e.g., a faster flow with greater density of information~\cite{heimgartner2007towards} and a lower level of driving anger~\cite{liu2013driving}. 


 
With globalization, the taxonomic view of Hofstede's cultural dimensions however gradually loses its original roots within geographical separation. Critiques of this model are widely pronounced in the field of technologically-enabled interactions, finding difficulties in applying the framework for explaining differences in technology use as well as in design~\cite{Ess2005,Marsden2008}. As a consequence, the concept of postcolonial computing is formulated not as a new domain or design space, but as an alternative sensitivity to the process of design and analysis~\cite{Irani2010}. It raises a series of questions and concerns inspired by the context of post-colonialism, relevant to any design project -- for example, but not limited, to HCI4D\footnote{see, for example, http://prior.sigchi.org/communities/hci4d} applications.
It prompts designers to consider their work as transformative interventions, abreast the progression from cultural nationalism towards a global village~\cite{Irani2010,Poll2012}. With that we also face the question whether cultural adaptation is sensible in a world in which users might deliberately acquire a German or Chinese car because they expect different experiences from them based on historical evidence.
Our approach to this work thus discards the idea of designing culturally specific interfaces in favor of letting users experience designs based on the needs of a culturally diverse user sample, in order to find disparities and common grounds.



\begin{figure*}
  \centering
  \includegraphics[width=\textwidth]{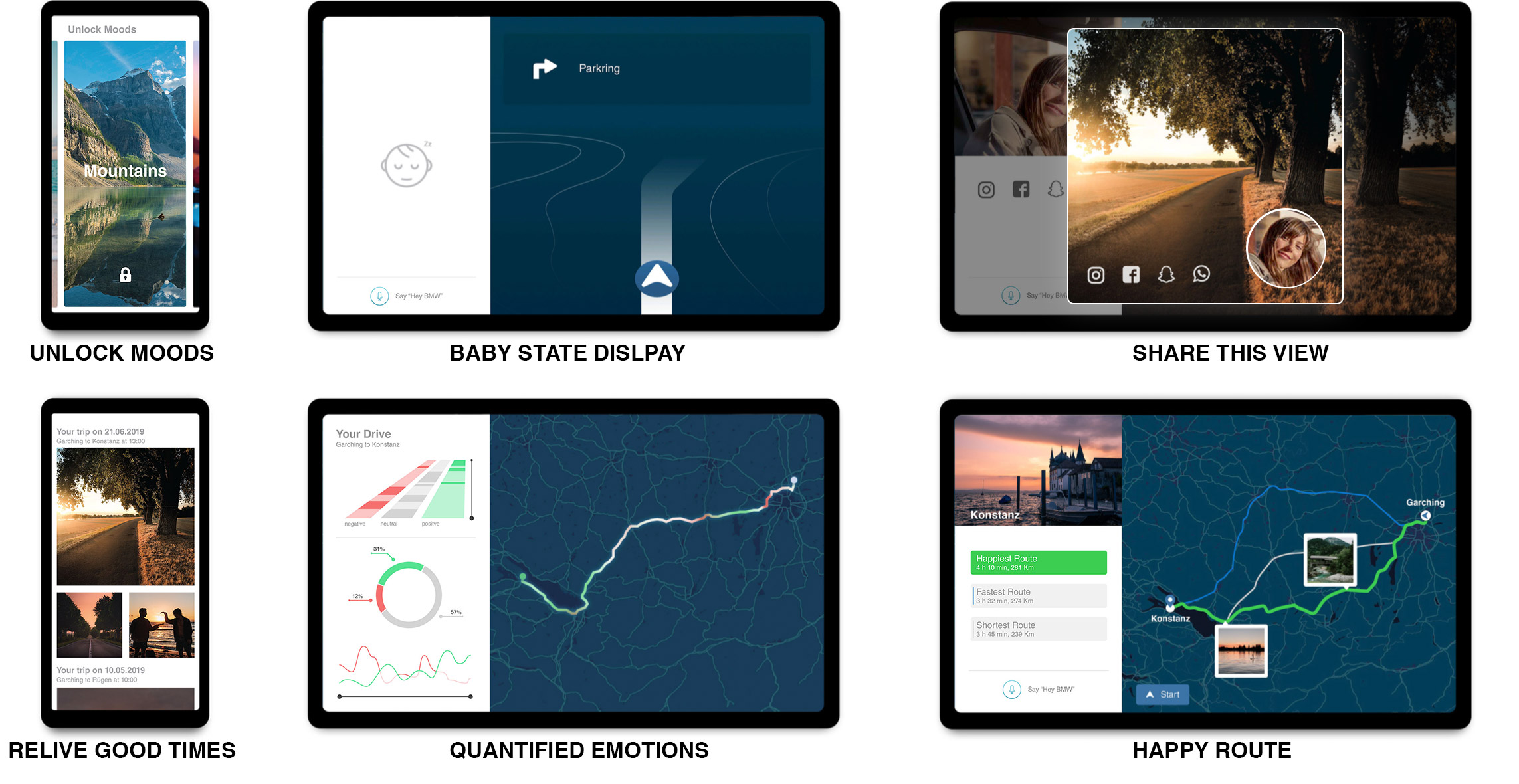}
  \caption{Some features are incorporated in a smartphone app, as they are interacted with before or after driving. Others offer a speech interface and GUI on the central information display. Depicted is one exemplary use case for each cluster. 
}~\label{fig:screens}
\end{figure*}

\section{Use Cases for Affective Interaction}

The design process we used to generate a culturally inclusive set of affective use cases was based on an inter-cultural design thinking approach previously used by Li et al.~\cite{Li2019}. We first collected a wide set of ideas from an analysis of related work and the output of 4 ideation sessions with a total of 50 German users who work in the automotive and HCI fields. We screened out unrealistic ideas and clustered similar concepts to provide essential insights for the following ideation session with Chinese participants. In the second phase, 13 Chinese users participated in pairwise car-storming\footnote{brainstorming in cars} sessions, ideating on applications specific for the Chinese market and also iterating on the previous ideas based on their experiences from driving in China. All participants showed interests in pragmatical use cases, such as navigation and empathetic voice assistants. However, Chinese users specifically took into account the family trip, such as driving with a baby or the elderly, since a newly formed family is the prevalent motivation for Chinese drivers purchasing new cars.

The resulting 50 use case ideas were then discussed in a workshop with eight experts on automotive user interfaces. Our intention is not to select uses cases that completely envelop the design space of affective in-car interaction. Rather, we aim to find the use cases that can be realistically executable in the car for users to interact for multiple occasions in different situations over longer time. The prioritization was done using a matrix with the dimensions ``potential'' to increase experience and ``feasibility''. We finally adopted all ideal use cases, i.e., those that were evaluated as both feasible and promising, and attentively kept the use cases which fulfilled only one criteria, while marked widely as favorites or merit further consideration, such as Smile to Unlock (security), Exterior Mood (privacy) and Passenger State Display (social desirability, paternalism). The final selection consisted of 20 use cases grouped into six thematic clusters. 

\subsection{Navigation Use Cases}

Affective data from the crowd can be combined with GPS information to enhance existing navigation functionalities~\cite{Pfleging2014}: In use case \usecase{Happy Route} the system not only offers the fastest and shortest routes but also a course which is based on the most positive emotions other drivers had along the journey. The feature \usecase{Avoid Frustration} offers the functionality to reroute the car when single spots on the trajectory cause very negative emotions in other cars in front of the driver, e.g., sudden impairments in sight or road damages. \usecase{Pleasant Parking} offers frustration-free parking spots based on the affective data of other drivers who have previously parked in the region. 


\subsection{Proactivity Use Cases}

These use cases are offered proactively by the voice assistant, reacting to the detected emotions. \usecase{Relive Good Times} can collect impressions from the ride and send them to the driver's phone after a trip. \usecase{Happy Place} offers sites on the way where other drivers showed positive emotions, whereas \usecase{Avoid Monotony} offers to entertain the driver on routes that other drivers perceived as boring. In case of bad mood inside the car, the proactive assistant switches to \usecase{Do Not Disturb} mode, which means that it will refrain from speech to not bother the driver and passengers even more.

\vspace{-0.1cm}

\subsection{Family Use Cases}

Focused towards young families, the function \usecase{Baby State Display} shows the detected state of the baby in the backseat. This was a direct requirement from Chinese participants. \usecase{Kids Games} offers games based on facial gestures as rear seat entertainment for children~\cite{mueller18smile}.


\subsection{Control Use Cases} 

This cluster contains functions which allow controls based on affective data: \usecase{Smile To Unlock} identifies the owner through face detection and opens the car when activated by a smile. 
\usecase{Unlock Moods} provides an interface to set a mood inside the car, which can be realized by adapting seats, lights, air conditioning, suspension, smells, and music. Users can rate the music they are listening to with facial expressions in the use case \usecase{Smile To Like}. The function \usecase{Emotional Playlists} further provides music selections based on songs they previously sang along to or that they tend to listen when they are sad or happy.

\subsection{Sharing Use Cases}

Users can activate the camera with a voice command and take a \usecase{Selfie} of themselves by smiling, which can then be shared with friends. The use case \usecase{Share The View} allows to take an additional picture of the outside scenery and share it via social media. With \usecase{Exterior Mood}, drivers can also communicate their emotional state to the other road users with a colored indication on the windshield visible from the outside.

\subsection{Transparency Use Cases}

Use cases in this cluster are designed to improve the transparency of the emotion detection system. Inspired by related work~\cite{braun2019driverstate,Voelkel2018}, \usecase{Driver State Display} and \usecase{Passenger State Display} show the detected states of the driver / all occupants as color-coded contour lights and in \usecase{Quantified View}, collected data is visualized as infographics. \usecase{Data Opt-in} finally lets users decide whether they would like to share their personal data to enable the affective functionalities described above.

We at last iterated on the speech interface and GUI design of 20 use cases from wireframes to the final interactive prototypes on smartphones and tablets (see Figure~\ref{fig:screens} for examples). 


\section{User Study}

During ideation sessions we were in contact with users and professionals who shared their experiences with technology in general and their expectations for emotion-aware systems in particular.
Drawing on this input and insights from related work, we derived a set of hypotheses to be tested in a study with German and Chinese drivers.
The null hypothesis assumed no differences for the pairwise comparison of use case demand and user experience between Chinese and German users. It will be rejected for certain use cases (cf. \Cref{fig:graph}).


Based on the potential improvements of user experience, which affective features have been shown to provide to drivers~\cite{Coughlin2011,Harris2011,Hassib2019}, we expected an overall high demand for affective UIs across cultures~\emph{(H1)}. We further expected emotional interaction to provide improvements on a hedonic, rather than on a pragmatic level~\emph{(H2)}, as emotional reactions are part of the definition of hedonic quality~\cite{Hassenzahl2005}.
Based on the contrasting cultural profile for the dimension individualism~\cite{hofstede2010}, we anticipated Chinese drivers to long for more socially supportive applications than Germans~\emph{(H3)}. Furthermore, a higher rating in uncertainty avoidance for Germans lead us to believe that German users would be more hesitant to trust and share data than Chinese users~\emph{(H4)}.
We  assumed that proactive features would be considered more paternalizing by Germans~\emph{(H5)}, as they tend to see suggestions as restrictive due to their low scores in power distance.


Related work also indicated that Chinese drivers are more likely to be open towards a functionality solely because it is new, while Germans prioritize practicability~\emph{(H6)}~\cite{Lachner2015}.
From our focus groups we learnt to expect 
that participants who stress the importance of transparency are often sceptic towards sharing personal data~\emph{(H7)} and that users are more open towards sharing their personal emotion detection data after they experienced a demonstration of its benefits~\emph{(H8)}.

\subsection{Study Design}

The study used a within-subject design in which each participant experienced all 20 use cases for emotion detection. The order of the use cases was counterbalanced using a latin square to avoid sequence effects. Only \usecase{Data Opt-in} was always at the final position, as participants were required to have experienced all previous ideas to evaluate it. This use case is thus reported separately.
Participants provided demographic data such as age and gender, the type of car they owned and mainly used, self-assessments on technical affinity and cultural allegiance, previous experiences with digital agents, voice interfaces and emotion detection, and their self-reported disposition towards empathetic and proactive digital assistants.

Each use case was evaluated using an 8-item UX questionnaire (UEQ-S, \cite{Schrepp2017}) and one-item trust and paternalism scales, accompanied by a short semi-structured interview\footnote{\label{interview}Interview structure: Would you want this function in a future car? Why (not)? In what situation would you use this functionality?}, after they had experienced the application. In the end, participants ranked the use case clusters according to their preferences and had the chance to discuss their most and least favorite functionalities. 



\subsection{Participants}

The study was conducted in [anon. city], China and [anon. city], Germany. For each country we recruited owners of premium cars\footnote{Sampled drivers owned cars by Audi, BMW, Cadillac, Infiniti, Land Rover, Lexus, Maserati, Mercedes, Mini, Porsche, Tesla and/or Volvo} aged 25 to 50 years, because young to middle-aged premium drivers are most likely to be the first adopters of innovations in cars and are, thus, our primary target group~\cite{rogers2010diffusion}. We made sure participants had been raised in the respective country and had not lived in another cultural region for longer than one year. This was important to us as we were interested in cultural differences, which might be influenced by extensive experience within other cultures. Participants were compensated for the approximately 90 minutes in the experiment with \euro\,75 or \yen\,800, respectively.

The 32 German participants 
were $39.4\pm7$ years old, 
16 self-identified as female and 16 as male. 
Fourteen participants had children under the age of 16 and only two were previously in contact with emotion detection, but in other contexts (art installation, work safety).
The Chinese sample consisted of 33 participants aged $36.8\pm4.6$ years, 16 of which self-identified as female and 17 as male. Twentytwo participants had children under the age of 16 and none of them had previous experiences with emotion detection technology.
All participants considered themselves technology-savvy\footnote{Based on a 4-item self-report questionnaire on technical affinity} 
and a vast majority stated they identified well with the culture that was publicly lived in their respective home country. 
Driving experience was well dispersed with 14\% of occasional drivers (<10,000 km/year), 56\% of moderately frequent drivers (10,000--20,000 km/year) and 31\% of frequent drivers (>20,000 km/year). 


\subsection{Apparatus}

\begin{figure}
  \centering
  \includegraphics[width=\columnwidth]{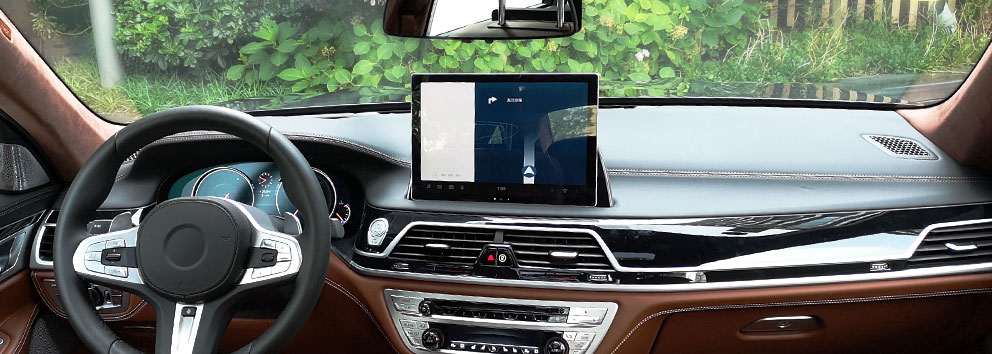}
  \caption{Prototype mounted in front of the CID. The system detects emotional states using a camera in the bottom frame of the device.}~\label{fig:apparatus}
\end{figure}

We installed the prototype in the cockpit of a premium class car. It was realized using Unity 3D, running on a Microsoft surface tablet mounted in front of the car's central information display (CID) and a 6.3'' Android phone. It included a GUI for all 20 use cases, which were activated and controlled by voice interaction. We simulated speech detection through a Wizard-of-Oz interface that was not apparent for users. For emotion detection we used the Affdex Emotion Detection SDK~\cite{McDuff2016}. This was fully functional, using the built-in camera positioned at the lower frame of the tablet, which provides an optimal field of view to the driver seat when mounted on top of the slightly tilted CID. The prototype was localized in German and Chinese language. \Cref{fig:apparatus} depicts the study setup. 
The tablet was also used to fill in the questionnaires after each use case, a GoPro camera recorded the experiment to re-evaluate participants' statements. 

\subsection{Procedure}

\begin{figure*}
  \centering
  \includegraphics[width=\textwidth]{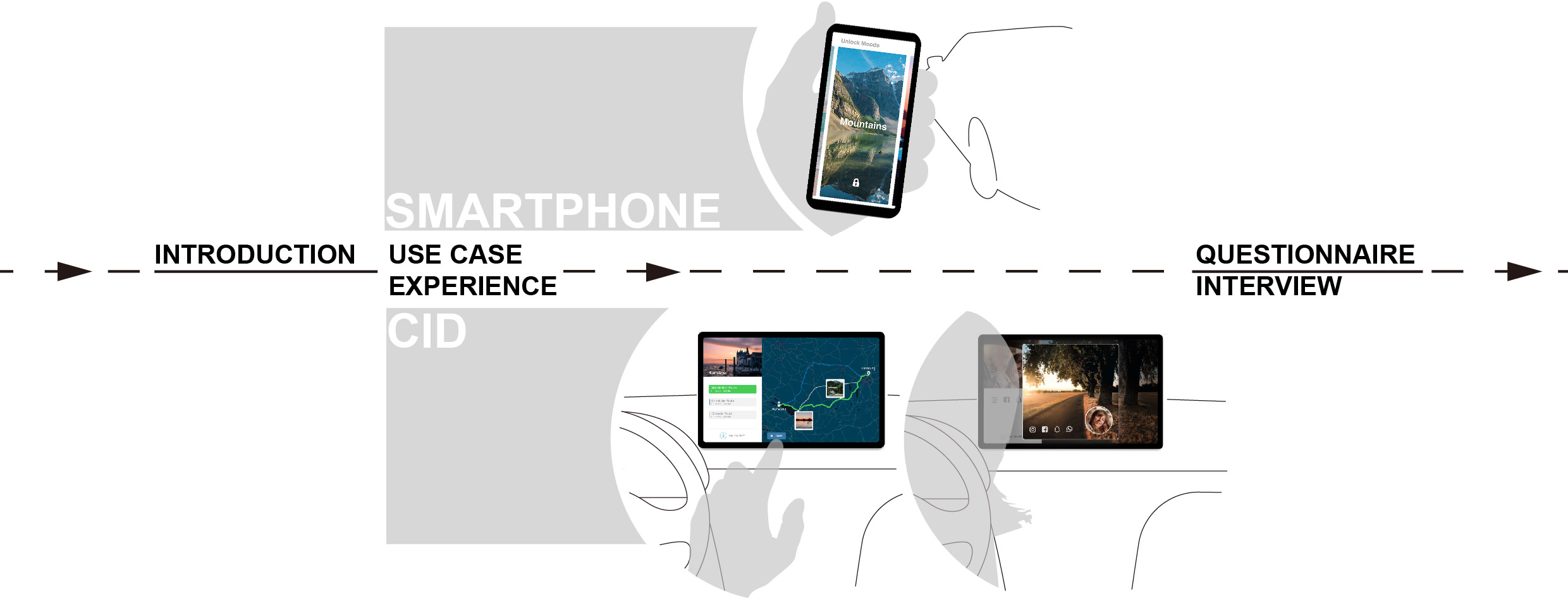}
  \caption{Study timeline for each use case experienced on the car's CID or outside the car on the smartphone.}~\label{fig:journey}
\end{figure*}

We invited participants to the research center and shortly introduced them to the topic of the experiment. Once the participants had taken a seat in the car they experienced a demo of the emotion detection algorithm in the form of a smiley reenacting the user's facial expressions. After playing around with the demo for some time, participants were informed that they will experience 20 use cases for emotion detection in a prototype and that they can interact with the system via touch and speech.
They then had the chance to read through the questionnaire and settle any open questions.
The experimenters also made sure to get across that they were not involved in the design and that they were open for both positive and negative feedback. 
They were further instructed to rather push for negative answers in the semi-structured interview when they assumed doubts about the feature.

We demonstrated all twenty use cases in the standing car on a secluded parking spot with natural lighting. \Cref{fig:journey} shows the presentation of each use case, which started with a short introduction by the experimenter, explaining the basic idea behind the function. Participants would then experience the functionality by interacting with the system, mostly through speech input and facial expressions. After each use case they filled in the 10-item questionnaire and conversed with the experimenter along a pre-defined interview guideline\textsuperscript{3}. 

The experimenters were local native speakers, conducting  
the study in the local language. The recruiter collected demographics in advance and forwarded it in an anonymous format. We immediately anonymized the recorded video files after the study to adhere to the strict data privacy laws of both countries.

\subsection{Limitations}

Participants were recruited among the owners of premium cars. This was a choice we made based on the fact that new technologies most likely first find their way into premium cars. This, however, also means that our findings can not necessarily be generalized to the entirety of the automotive market.
Furthermore, the user sample in China was younger and drove more luxurious cars than the sample in Germany, which at a first glance might seem imbalanced but represents the market segmentation in these two countries quite well~\cite{schrott2014strategies}.


\begin{figure*}[t]
  \centering
  \includegraphics[width=1.065\textwidth]{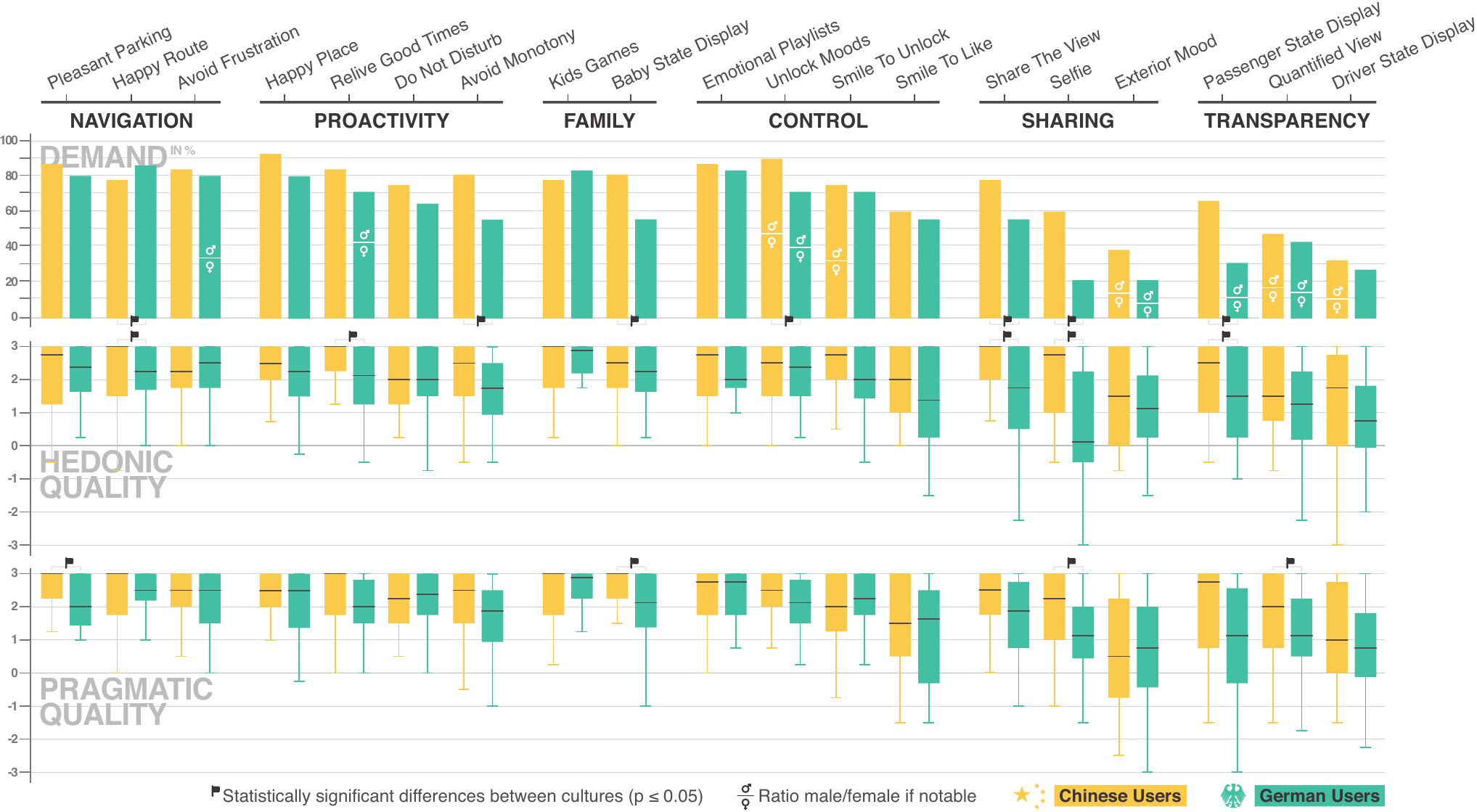} 
  \caption{Demand, hedonic and pragmatic quality for the tested use cases, sorted by most liked clusters. The numbers for demand -- meaning how many participants stated they want this functionality in a future car -- show
  tendencies comparable to the UX measures hedonic and pragmatic quality. Countries are color coded, gender differences are indicated within the demand columns if noteworthy.}~\label{fig:graph}
\end{figure*}


\section{Results}
We compared quantitative UX measures split by cultural backgrounds and gender. As all participants self-identified as either Chinese or German and male or female, we used binary differentiation for pairwise comparison (t-test in JASP~\cite{jasp}). We analyzed demographic data and use case preferences based on Pearson correlation coefficients. We report statistical significance for $p \leq 0.05$, correlation effects are reported as weak/moderate/strong with $r \geq 0.1/0.3/0.5$~\cite{correlations}. Feedback from the interviews is made accessible through thematic analysis~\cite{Braun2006}, key themes are underlined using verbatim quotes.


\subsection{Demand}
Overall feedback from the post-experiment interview identified the use case cluster \navi{} as the most desired in both countries. \cont{} use cases are placed second in Germany and third in China, \fami{} use cases are conversely ranked second in China and third in Germany. \proa{} use cases were ranked fourth in both countries, followed by the least desired clusters \shar{} and \tran{}. In general, more than two thirds of the participants expressed the demand for affective automotive interaction after the experience.

The top part of \Cref{fig:graph} shows the demand for each use case as stated by users in interviews after each interaction. \navi{} use cases are most sought after and the majority of \proa, \fami{} and \cont{} functionalities are also perceived quite positively. The features \usecase{Relive Good Times} and \usecase{Unlock Moods}, which provide emotional experiences, are preferred by women, whereas the expressive use case \usecase{Exterior Mood} attracts more interest from men.
The feedback of Germans is often more sceptical compared to Chinese users, especially when it comes to \shar{} functionalities. \tran{} use cases are generally disliked except for \usecase{Passenger State Display} which many Chinese participants find useful. Features from this cluster are overall more wanted by men than by women.


\subsection{User Experience}

Subjective measures of user experience follow a trend comparable to the tendencies for demand. Differences in hedonic and pragmatic aspects go hand in hand with a discrepancy in demand between cultures (see \Cref{fig:graph}). The most disparate ratings were recorded for the use case \usecase{Selfie}, with high UX assessments by Chinese users and comparatively negative evaluations by Germans. The feature \usecase{Unlock Moods} received higher ratings for its hedonic quality by women than by men, \usecase{Exterior Mood} and \usecase{Passenger State Display} were rated more pragmatic by men than by women and \usecase{Quantified View} is rated higher in both categories by men. It is striking that none of the use cases show blatantly negative ratings, which raises the suspicion of a positive bias due to the novelty of the technology or a potentially high acceptance towards affective features across the sample. 



\subsection{Trust \& Paternalism}
With increasing automation and the advent of proactive assistants, trust and paternalism are becoming important performance indicators for user interfaces. \Cref{tab:means} shows the mean values for all use cases without a differentiation between cultures as there are only minuscule differences. Germans rated \usecase{Pleasant Parking} as less trustworthy and \usecase{Emotional Playlists} as more paternalizing, while Chinese saw \usecase{Smile To Unlock} as more paternalizing. There were no differences between genders. Most ratings for trust are in the rather vacuous lower to moderately positive ranges, most likely because no feature accessed safety-critical functionalities. Paternalism scores were notably high for the use cases \usecase{Exterior Mood} and \usecase{Passenger State Display}.


\begin{table*}[htb]
  \includegraphics[width=\textwidth]{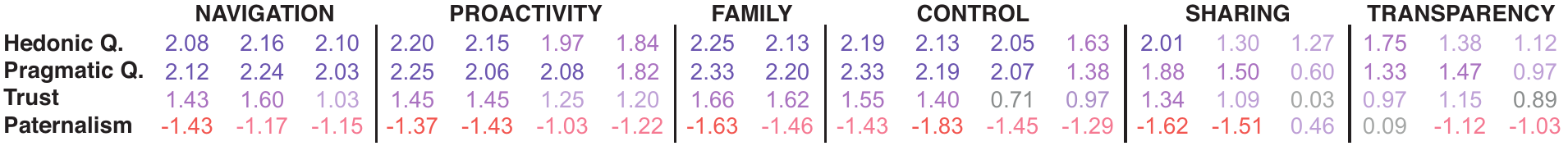}
\caption{Overall mean values for user experience qualities, trust and paternalism rated on a scale from -3 (red) to +3 (blue), order as in \Cref{fig:graph}.}
\label{tab:means}
\end{table*}


\subsection{Data Privacy}

Participants were asked twice whether they would share their affective data with [anon company] if it added value to the in-car UI. Before the experiment, 79\% of the Chinese and 84\% of the German users said they would. After experiencing all use cases, 82\% of the Chinese and 91\% of the German participants stated a willingness to share their data to enable the presented features.

\subsection{Demographic Influences}

Considering the demographic backgrounds of participants, we analyzed relations among the data collected in the pre-questionnaire.  
The data shows that positive assessment of the \fami{} cluster moderately correlates with the user's age ($r = 0.352$), which makes sense as increasing age is connected with a higher likelihood of having children ($r = 0.188$). Interestingly, users with larger families tend to like these features less ($r = -0.322$). The latter can be explained by the facts that Chinese participants ranked this cluster higher than Germans and that they, on average, have less children per family (1.04 vs.~1.69). 
Furthermore, we found that users who stated they felt paternalized more often did not want an interface to assess their emotional state ($r = 0.380$) and also hesitated to accept proactive system behavior ($r = 0.466$). Acceptance of emotion recognition and proactivity were also moderately correlated  ($r = 0.346$).
Finally, participants' rankings for the use case clusters \shar{} and \tran{} are inversely correlated ($r = -0.398$), which seems obvious but also suggests divided opinions concerning the usage of private data.


\begin{figure}[t]
  \centering
  \includegraphics[width=\columnwidth]{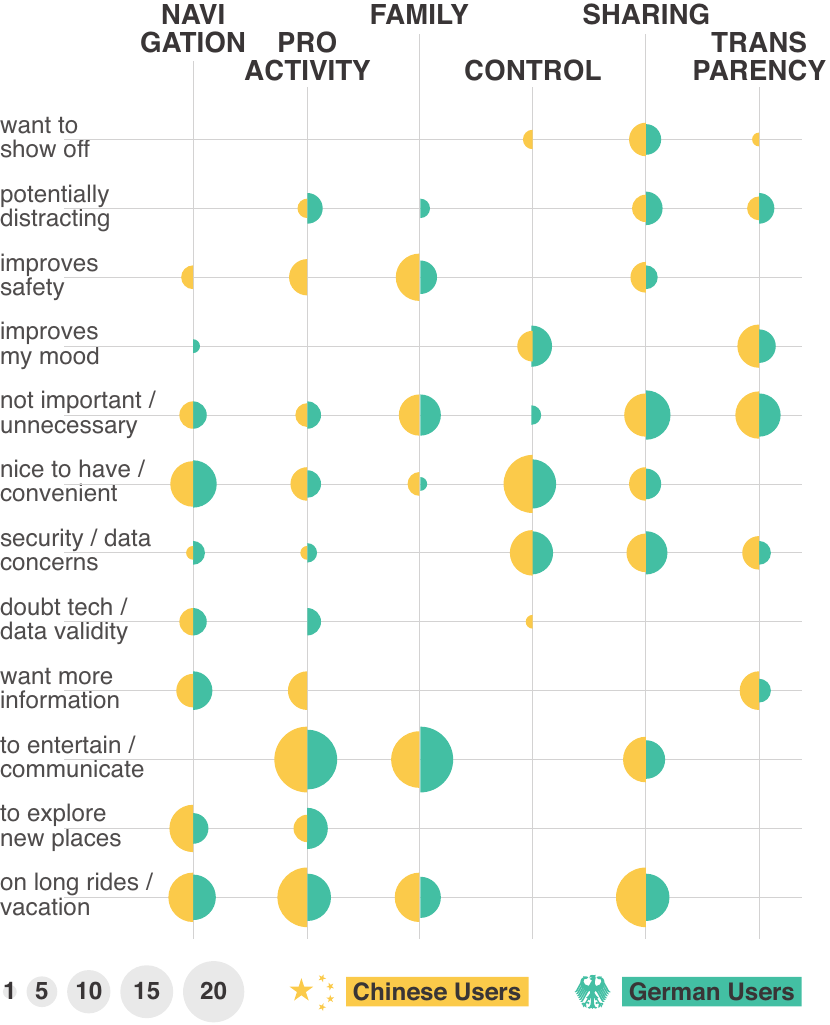}
  \caption{Main themes from the qualitative feedback grouped by use case clusters and cultural background.}~\label{fig:interview}
\end{figure}

\section{Qualitative Feedback}

Participants were asked to think aloud while experiencing each use case and afterwards share their assessment with the examiner. 
The German and Chinese experimenters worked together to develop a set of recurring themes in English language, using thematic analysis on the original notes and recordings~\cite{Braun2006}. The resulting themes are visualized in \Cref{fig:interview}, direct quotes are translated as closely as possible and identified with user IDs.

\subsection{Navigation Use Cases}

Participants in both countries liked \navi{} features the most, as they are convenient additions to common functionalities. \usecase{Pleasant Parking} was the overall best rated use case and many users seemed to share the view that \quot{...looking for parking spots is very frustrating, so reducing [their] frustration is an added value.}{G16}. Many, however, asked for more pragmatic information \quot{...such as distance and remaining parking spots and facilities}{C7} or \quot{price information}{C16}, showing that no benefit of emotion detection as a data provider for a parking scenario was perceived.
In contrast, emotion-based routing as shown in \usecase{Happy Route} was assessed as a great way to improve the quality of the ride. One user stated \quot{I often take longer routes If I think it makes me happy.}{G22} although it is seen as a \quot{...huge effort to look for the nicest route and avoid frustrating parts by myself.}{G27}. The features are also appreciated as being unobtrusive: \quot{It's a suggestion and I can choose it or not, that's pleasant.}{G8}. Most users said they would use this to explore unknown regions, e.g., on vacation. Chinese users especially related the function more often with joyful family trips.

\subsection{Proactivity Use Cases}

Many participants seemed to welcome the idea of proactive interaction as a means for simplification: \quot{Everything the car does, I don't have to do.}{G13}. \usecase{Happy Place} provided an obvious benefit on long journeys, as many drivers saw service stations as sub-par but without alternative when in need of a rest: \quot{This is great. I think I often pass places I don't know about, where I could have made a break with a great view.}{G16}
Proactivity can in the mind of many help to improve driver safety by taking off some work, others, however, also see a potential for distraction: \quot{What if the assistant distracts me and I cause an accident?}{G11}.
The use case \usecase{Avoid Monotony} is mostly seen as simple entertainment, which some users find unnecessary as they say it is fine to \quot{ ...just get bored for a moment. No need to be happy all the time on the road.}{C8}. At the same time more Chinese users found it beneficial in long-distance driving. \usecase{Relive Good Times} was liked well by Chinese users, who saw it as spadework for sharing (\quot{Automatic photos make it easier to share with friends in WeChat in time.}{C22}, \quot{...perfect for travels with my child and dog. The album after the ride can be shared on TikTok.}{C28}). Among German users, women liked the memories created but many men did not see the benefit.

\subsection{Family Use Cases}

Use cases centred around children are widely accepted, however only for small children. German parents, who on average have more children and are thus supposedly more experienced, were rather sceptical towards the \usecase{Baby State Display}, as they said one \quot{...can hear perfectly fine how the baby is feeling. It is communicating all the time}{G10}. Users also said they would prefer a direct camera stream instead of icons, so they could react if the child intended to \quot{open the door}{C33}, \quot{put something in its mouth}{C14} or \quot{unlock the safety belt}{C20}.

\subsection{Control Use Cases}

These features were mostly seen as \quot{funny kick-knacks.}{G28} which could improve the mood of the driver because they are fun to use. They were assessed positively but also seen as superfluous, e.g., \usecase{Emotional Playlists} was described as \quot{not an important function, but I still want it.}{C8}.
\usecase{Smile To Unlock} was on the one hand praised for being more convenient than accessing the car with keys or finger prints. On the other hand users were displeased with possible security issues and the need to fake a smile which mainly Chinese users found patronizing.
A minority of users found that it inevitably improved their mood when \quot{every day starts with a smile}{G30}.
The feature \usecase{Unlock Moods} was more popular with women than men and most users wished for the possibility to customize their experience instead of being given pre-set choices. Moreover, it was valued by Chinese participants \quot{to emotionally get myself ready with following tasks}{C14}.

\subsection{Sharing Use Cases}

Many Chinese users showed a passion for \shar{} features. Especially the \usecase{Selfie} use case seemed already quite common among Chinese drivers: \quot{I like selfies so much! Usually I take selfies at every red light}{C13}, \quot{I like taking selfies to record each moment of my emotion. I usually do it when I stop at a red light and feel bored}{C19}. They also appreciated the potential usage of a group in-car selfie. Among German users, many stated a feeling of not being part of \quot{generation selfie}{G10}, \quot{Nobody needs this, or I'm just too old.}{G28}. Younger drivers from Germany also liked the feature: \quot{I want to send a selfie when I start. Right after preparing my outfit I always look the best}{G7}.
The use case \usecase{Share The View} was rated better because it added value in a social context: \quot{I'm not the selfie type but I could use it to keep my family informed}{G8} and in professional situations: \quot{I can update my boss who asked where I am driving now}{C20}.
Overall, \shar{} features were seen by many as not important but nice to have (\quot{I don't see the sense but I like it}{G11}). Chinese participants were more open to accept them than Germans. They saw a benefit in having show-off features they could impress others with and the pragmatic value of keeping in contact with group members, however users from both countries also expressed concerns regarding data privacy and distraction.

\subsection{Transparency Use Cases}

The transparent visualization of collected data was assessed as the least desired use case cluster. Participants described the features to be interesting once or twice but not important in the long run. There was a slight tendency for men to be more open towards \tran{} use cases.
The \usecase{Driver State Display} was penalized as being too intrusive and unprofessional. 
Likewise, the use case \usecase{Quantified View} was seen as an unnecessary presentation for drivers, but should rather serve the purpose of background processing to improve big data systems for some Chinese users.
Only the use case \usecase{Passenger State Display} was accepted by Chinese users, who can imagine \quot{to take care of other passengers}{C9} based on the system feedback. They feel that \quot{as the car serves as a transporting space shared by a number of people, [the driver] should also coordinate between them}{C28}.


\section{Implications for Affective In-Car Interaction} 

The feedback from both groups confirms a high demand for affective automotive UIs across cultures \emph{(H1)}. Especially use cases which use emotions to improve well-tried interactions or improve comfort through proactive behavior were rated favorably by many participants. Furthermore, we found that cultural and social factors impact what users accept and expect from affective interfaces and that participants showed active awareness for the privacy of the gathered data. These main insights allow us to formulate implications for future HCI research on emotional in-car interaction:

\subsection{Emotion-Awareness Improves Established Features}

We initially assumed that many affective use cases would be perceived as purely hedonic features, as emotions are hardly a pragmatic construct~\emph{(H2)}. The demand data, however, showed that primarily pragmatic features, such as navigation and parking, were desired applications for affective systems. \Cref{fig:graph} shows that for a majority of use cases, pragmatic quality was rated comparably high or higher than hedonic quality. Participants see the emotional aspect of these use cases as convenient additions to common functionalities they already know. This also holds true for proactive features, which add a layer of system-initiated interactions to established functionalities. 

Such an emotional upgrade, however, also introduces new challenges, especially when driving related and safety critical features are enhanced.
We reason from these insights, that an augmentation of established features with emotional interaction could be a promising strategy for the near-term introduction of affective interaction, paving the way for more advanced applications in the upcoming future. 

\subsection{Proactivity Does Not Necessitate Paternalism}

In contrast to our initial assumption that proactive features might be perceived as paternalistic, especially by German users (\emph{H5}), only the use cases \usecase{Exterior Mood} and \usecase{Passenger State Display} were evaluated as rather patronizing (see \Cref{tab:means}).
Both cultural groups are less distinct than anticipated when it comes to perceived paternalism, as Germans only assess the use case \usecase{Emotional Playlists} more paternalizing than Chinese users. The Chinese in contrast perceive \usecase{Smile To Unlock} as more paternalizing, which can be explained with the advanced digitization prevalent in the Chinese society: as Chinese citizens are used to payment methods and public surveillance using facial recognition, they are biased by existing experiences associated with this technology.

This makes us believe that not proactive system behavior in itself, as we assumed initially, but rather the underlying assumptions regarding driver and passenger states and the unfiltered communication thereof, have strong effects on how paternalistic a system is perceived. When designing such user interfaces we consequently need to think of the context the interaction takes place in and whether the intended communication of data is appropriate.



\vspace{0.6cm}

\subsection{Losing Control Over Private Data Is A Dealbreaker} 

While we could confirm the expected positive effect of experiencing comprehensible use cases on the willingness to share private data in both countries~\emph{(H8)}, we also have to object to the presumption that Germans are more hesitant than Chinese to trust the system~\emph{(H4)}. In fact, Chinese users showed more reservations against sharing their emotion detection data with the provider than Germans (82\% vs. 91\%). We argue that there is a change in attitude towards privacy in China with increasing surveillance, while the notoriously private Germans more and more accept a post-privacy stance, previously exemplified by other western countries like the United States.

Interestingly, the ratings for use cases of the cluster \tran{} are inversely correlated to \shar{} features, which substantiates the division into a partition of data-aware users and a rather sharing-driven group within the whole sample~\emph{(H7)}. In this rather mindset than regionally dependent finding we also discover arguments for a post-colonial approach, entailing an adaptation based on a user's beliefs instead of cultural affiliation.

\vspace{-0.1cm}

\subsection{Personal Choice Comes Before Cultural Adaptation}

Concerning the differences in demand between cultures, we saw that Chinese users were indeed more enthused by the proposed features than Germans, allowing us to confirm~\emph{H6}. The qualitative feedback, however, shows that the notion of wanting to show off novel features is also prevalent in the German sample (see \Cref{fig:interview}). We further see differences between genders, as female participants were to some extent more open towards emotionalizing the driving experience, e.g., through setting a mood inside the car with the feature \usecase{Unlock Moods}.
Based on these insights, we will have a hard time offering users the right features based on their cultural background, without knowing more about their needs and interests first. One approach to raise acceptance of affective UIs could thus be to let customers experience a wide set of possible trial offers and let them chose which ones to keep.


\subsection{Affective Features Need Socio-Cultural Responsiveness}

Chinese users showed significantly higher demand for the use cases \usecase{Passenger State Display}, \usecase{Selfie} and \usecase{Share The View}, compared to German participants.
Chinese drivers are also more inclined to share personal data to keep connected with social networks. This allows us to accept our hypothesis that social interaction is of high significance in China~\emph{(H3)}. The feature \usecase{Passenger State Display} enables them to take care of other passengers and fulfill their  responsibility as the coordinator in the car. They cherish the features \usecase{Selfie} and \usecase{Share The View} as they allow them to share their experiences and show their belonging to a group or family, and impress others to strengthen their social identity. 
Only 20\% of the German participants, however, approved of the use case \usecase{Selfie}, and \usecase{Passenger State Display} was also seen critically. 

This leads us back to the fact that individualism shows the main discrepancy between German and Chinese users within the cultural dimensions. In this context, an affective UI should offer users from rather collectivist societies more occasions for social experiences, while users from more individualist societies might perceive too much of them as overwhelming.


\section{Conclusion and Outlook}

In this work we explored a set of 20 use cases for affective automotive user interfaces to close the research gap between already quite advanced emotion detection technologies and still highly theoretical application concepts. Feedback from German and Chinese drivers confirms an overall demand for emotional automotive interaction and demonstrates that cultural differences are to be considered when designing such interfaces. Yet social influences and aspects of global developments are equally important.

We provide concrete starting points to realize affective automotive UIs by 1) improving established features with emotional components to gain awareness and create acceptance for affective interaction; 2) preventing paternalism in proactive scenarios through context sensitivity; 3) catering to recent cultural developments, such as privacy awareness in China and a post-privacy attitude in Germany; 4) leave room for personal choice instead of automatic adaptation; and 5) show cultural awareness for the big topics, e.g., the importance of social situations in different cultures.

We want to inspire both researchers and practitioners to apply these implications to their work on affective automotive user UIs, so we can soon benefit from emotional interaction in our cars. Future research can build upon our insights to improve the acceptance of their interfaces in different cultures. This could, for example, be interesting for work on natural user interfaces, like conversational UIs, as affective computing allows for adding empathic features or social awareness. Another interesting research area is the viability of cultural dimensions in a globalized consumer market, as we see arguments for the benefits of cultural adaptation but also effects which take issue with this regional approach. Finally, we see potential in researching privacy-enabled user interfaces like head-up displays, holograms, or private audio zones, which could minimize social pressure in the presence of other passengers and thus lead to more acceptance towards emotional interaction in the car.

\begin{acks}
We want to thank Guillermo Ponce Zambrana and Lea Grötzinger for their support regarding interface design and graphics.
\end{acks}

\bibliographystyle{ACM-Reference-Format}
\bibliography{bib}

\end{document}